\documentclass[conference]{IEEEtran}
\IEEEoverridecommandlockouts
\usepackage{tikz}
\usepackage{setspace}
\usepackage{enumerate}
\usepackage{cite}
\usepackage{times}
\usepackage{url}
\usepackage{graphicx}
\usepackage{subfigure}
\usepackage{amsmath}
\usepackage{algorithm}
\usepackage[noend]{algpseudocode}
\makeatletter
\def\BState{\State\hskip-\ALG@thistlm}
\makeatother

\begin{document}
\title{SP\_Async:Single Source Shortest Path in Asynchronous Mode on MPI}
\author{
\IEEEauthorblockN{Sangeeta Yadav\IEEEauthorrefmark{1},
Asif Khan\IEEEauthorrefmark{2}}
\IEEEauthorblockA{Department of Computational and Data Science,
Indian Institute of Science\\
Bangalore,India\\
Email: \IEEEauthorrefmark{1}sangeetay@iisc.ac.in,
\IEEEauthorrefmark{2}asifkhan@iisc.ac.in}}

\maketitle
\begin{abstract}
Finding single source shortest path is a very ubiquitous problem. But with the increasing size of large datasets in important application like social network data-mining, network topology determination-efficient parallelization of these techniques is needed to match the need of really large graphs. We present a new Inter node-bellman cum Intra node Dijkstra technique implemented in MPI to solve SSSP problem. We have used a triangle based edge pruning for idle processes, and two different techniques for termination detection. Within each node the algorithm works as Dijkstra and for outer communication it behaves as inter node bellman ford. First termination detection technique is based on the token ring and counter. Second is a heuristic based technique, in which the timeout is calculated from the number of inter-edges and number of partitions. In this project asynchronous mode of message passing is used.

\end{abstract}

\section{Introduction}
\label{intro}
In the linear programming formulation the SSSP problem can be defined as follows.Here we are given a directed weighted graph G(V,E) with source node S ,target node t, and a cost set W
\newline
we have to minimize $ \sum_{ij\in\mathit{A}} {w_{ij}x_{ij}} $ \newline
subject to $ x \geq 0 $ and for all i,\newline

$ \sum _j x_{ij}-\sum_j x_{ji} = \left\{\begin{array}{@{}ll@{}rr@{}}
    1, & \text{if}\ i = S \\
    -1, & \text{if}\ i = t \\
    0 , & \text{otherwise} \end{array}\right. $  
    \newline
\newline
\newline
SSSP is used as kernel and on its own for many real world applications ranging from data analysis, Internet, biological networks, centrality measures in social networks. The days are gone when an algorithm like Dijkstra was sufficient to solve SSSP for smaller graphs. Now the size is in billions, and thus sequential algorithm will take very long time to compute SSSP for these graphs on single machine. Many parallel algorithms/graph libraries and frameworks have already been proposed for finding single source shortest path to reduce this computation time. This paper details SP-Asyn (Single Source Shortest Path in asynchronous mode)  technique, which is parallelized form of inter node Dijkstra and intra node Bellman Ford algorithm. 
Various algorithm of SSSP in sequential and parallel form have already been proposed like DSMR \cite{maleki2016dsmr} and delta - stepping\cite{Chakaravarthy2017}. However all these works for synchronous mode only. Here we are proposing a unique technique in asynchronous mode. The major contributions of this project are:

\textbf{1.Trishla :} Triangle inequality based edge elimination technique which reduces the number of edges,and thus the unwanted edge relaxation steps(TEPS) within any process.

\textbf{2. SP-Async :} An asynchronous SSSP technique which combines intra-process Dijkstra with inter-process Bellman Ford algorithm.

\textbf{3.ToKa :} Two termination detection techniques have been proposed detailed as follows:
\newline \textbf{ToKa1:}This termination detection technique combines token ring algorithm and a counter to keep a check on the idleness of a process in the given graph.\newline
\textbf{ToKa2:} It is based on the timeout calculated by the product of number of graph partitions and the number of inter-edges among the processes.The idle process will wait for the incoming messages only within this timeout range. 
\newline
The paper is organized as follows Section II gives the literature review of the existing Parallel SSSP techniques, section III details the methodology adopted followed by the details of Experiments and Results given in Section IV.The project is concluded in Section V. Section VI tells about the future scope.
\section{Related Work}
Many parallel SSSP techniques have been developed.
Pregel \cite{Malewicz}, which is general vertex centric graph processing framework following bulk synchronous paradigm, solves the problem at the abstraction of each vertex. In every superstep a vertex updates its distance based on its previous value and received messages from vertices in previous superstep . If it changes its distance value then it sends update messages to all its neighboring vertices in the next superstep. Processes synchronize between supersteps. This approach does not work well when there is imbalance as fast processes are made to wait for synchronization because of a slow process.\newline \indent\indent\indent\indent Bellman-Ford based parallel algorithms do the same as in \cite{Malewicz}, except for synchronization. Between asynchronous and synchronous methods, although processes may have to wait for long but the work done is more efficient than in asynchronous methods, for synchronous. $\Delta$-Stepping \cite{meyer2003delta} tries to balance the two by using a parameter $\Delta$ which can be thought of as controlling degree of asynchronousness.
It uses a concept of distance buckets which contain vertices with distances in some delta size interval. For edge relaxations, the bucket with smallest index is chosen and all their vertices are relaxed. Processes keep their own buckets, but are synchronized before moving to process next lowest index non empty buffer. Keeping the $\Delta$ large allows for more vertices to be processed in one superstep and hence more asynchronicity but possibly work inefficiency. On the other hand small values limits the degree of asynchronicity and increases the number of required synchronization steps. In \cite{Chakaravarthy2017} authors improve upon the baseline $\Delta$-stepping by employing push and pull heuristics which is essentially deciding when to send updates to neighbors or ask for updates from neighbors. A Recent development, DSMR \cite{maleki2016dsmr} shows improvement over delta-stepping for a special class of graphs namely scale-free networks. Instead of distance discretization in delta stepping for synchronization they used a parameter D which limits the number of relaxations in the superstep. In each superstep every process employs Dijkstra based algorithm on its subgraph and for ghost vertices it buffers their updates through it. D limit is reached the ghost vertex updates are all to all exchanged and thus synchronized. They also use some preprocessing to eliminate redundant edges. We also use Dijkstra based algorithm for local computations but do not use synchronizations. Also we overlap redundant edge elimination with actual SSSP computations for some processes. Delta stepping and DSMR are current state of the art in distributed SSSP.

\section{Methodology}
\textbf{A. Graph Partitioning :}
Given graph is partitioned using 1D block partitioning. Knowing the total number of vertices N in the graph and the total number of available processes P, a partition id $ Pid = N/P$ will be provided to each vertex.To reduce the communication overhead incurred in graph distribution and adjacency list Padj will be maintained with each process such that $ Padj = \left\{\begin{array}{@{}ll@{}}
    Non-empty, & \text{if}\ v\in\mathit{P} \\
    empty , & \text{otherwise} \end{array}\right. $

\textbf{\newline B. Triangle based edge elimination:}
For process which do not receive any message they remain inactive and waste there cycles waiting for updates. In this section, we use this idle time to find and eliminate edges in the subgraph which will be guaranteed not to be included in any shortest path. We consider this as useful work since all the edges are considered for relaxation by the Dijkstra part and hence there removal will result in less work for Dijkstra.\newline
\indent\indent\indent We remove edges using Trishla in Algorithm1. We can see the correctness by arguing that we can always reach the other end of deleted edge indirectly by using the remaining edges in the triangle and that too with lesser path distance.
\begin{algorithm}
\caption{My algorithm}\label{euclid}
\begin{algorithmic}[1]
\Procedure{trishla}{}
\ForAll{$u \in\mathit{P_i}$}    
\ForAll{$\textit{$ v_i ,v_j \in\mathit{N(u)} and (v_i , v_j) \in\mathit{E_i} $ } $}
\If {$(w(u,v_j) > w(u, v_i ) + w( v_i,v_j ))  $} 
\State $delete (u_i,u_j)$  
\EndIf
\EndFor
\EndFor
\EndProcedure
\end{algorithmic}
\end{algorithm}
\newline
\newline
\newline
\textbf{C. Asynchronous SSSP:}
Our distributed Dijkstra algorithm is executed in parallel across the processes. Every process initially calls SP-Asynch with the source vertex as input, and afterwards it remains in this algorithm until termination. Line 5 is executed only by the process owning the source vertex. Initially only the source vertex owning process is executing the Dijkstra algorithm on its subgraph and if it encounters a vertex not in its partition during edge relaxations, it sends message to the corresponding owner process of the vertex. Other processes are executing edge elimination procedure and are activated upon receiving any message i.e switch to work on Dijkstra.\newline \indent\indent\indent\indent Essentially our algorithm is Dijkstra for subgraph at each process and Bellman-Ford at the level of graph of processes. When the process execution returns from one call to DIJKSTRA it calls TOKA termination detection routine assuming that it has no more work to do. If it ever receives work from some process at lines 9-13 it again calls DIJKSTRA and so on. When the TOKA routine detects termination, details of which are discussed next, all the processes break-out of the infinite while loop of line 8 and terminate.         
\begin{algorithm}
\caption{SP-Async}\label{euclid}
\begin{algorithmic}[1]
\BState \emph{Initialize}:
\State $d(v_i)  \gets \infty$.
\State $ \text{priority queue pq } \gets \emptyset$.
\BState \emph{Initialize}.
\If {$ s \in\mathit{P_i} $}
\State  $d(s) \gets 0 $.
\State $pq.push(s)$
\EndIf
\While{true}
\State $recv\_msgs ( anysource ,(v,d_v))$.
\ForAll{$(v,d_v)\in\mathit{msgs}$}
  \If{$d(v)> d_v$}
  \State $d(v) \gets d_v$
  \State $pq.push(v)$
  \EndIf
  \EndFor
  \State \text{DIJKSTRA(pq);}
  \State \text{TOKA();}
\EndWhile

\end{algorithmic}
\end{algorithm}
\begin{algorithm}
\caption{}\label{euclid}
\begin{algorithmic}[1]
\Procedure{dijkstra(queue)}{}
\While{ $\textit{pq is not empty}$}
\State $ \textit{u } \gets \textit{pq.pop()}$.
\ForAll{$v \in\mathit{N(u)}$}
\If{$(d(v) > d(u) +w(u,v) and v \in\mathit(P_i)$}
\State $ d(v) \gets d(u)+w(u,v)$
\State $ d(v) \gets pq.push(v)$
\ElsIf{$v \not\in\mathit P_i$}
\State $send\_msg(P(v),d(u)+w(u,v))$
\EndIf
\EndFor
\EndWhile
\EndProcedure
\end{algorithmic}
\end{algorithm}

\textbf{D. Termination Detection :}
In Bulk Synchronous mode every process synchronizes with the others before moving on to the next superstep, which ensures that every process knows when all others are idle and hence can terminate the execution. But for asynchronous case no process is sure if it has received all the messages intended to it and hence is not sure when to terminate. So we need some mechanism for the processes to consistently detect termination i.e all processes are idle. \newline For our purpose we use two methods. One is a heuristic based on the number of inter-edges among processes and one is token ring with message count based algorithm. In the first method every process maintains a \textit{msg\_count} variable which stores the number of messages received. If the \textit{msg\_count} becomes larger than number of border edges for the process times number of processes for the process, then it is terminated.\newline 
In the second method every process maintains a state variable which can be white, black or red and a count variable initialized to zero. It enforces some operations along with the msg send and msg receive operations sending or receiving distance updates. These operations are, set your state variable to \textit{black} and decrement your \textit{count} if sending and increment if receiving. Now, termination detection routine works by exchange of special tokens in a logical ring of processes with process $p_0$ being the initiator of token exchanges. Tokens contain a state and count. When process $p_0$ becomes idle it sends a token with its state and count values to next process in ring. All other process do the same, when they receive a token and become idle. If process $p_0$ receives a black or non-zero token it re initiates the token propagation otherwise if it receives \textit{white} and zero count token it sends a \textit{red} token into the ring and thus informing others of termination, and terminates itself. Termination related variables and their updation on actions discussed above are not explicitly shown in SP-Asynch, DIJKSTRA algorithm for readabilty.
\begin{algorithm}
\caption{Toka1}\label{euclid}
\begin{algorithmic}
\If{$msg\_count\geq numofPart*num\_of\_interedges$}
\State \text{terminate}
\EndIf
\end{algorithmic}
\end{algorithm}
\begin{algorithm}
\caption{Toka2}\label{euclid}
\begin{algorithmic}[1]
\BState \emph{Initialize}:
\State $state = white,count =0 $.
\BState \emph{Initialize}.
\State $recv\_mesgs(prev\_rank, \&state1,\&count1)$.
\State $state \gets \text{state OR state}1$.
\State $count \gets count+count1$.
\If{$myRank == 0$}
  \If{$state = white , count =0$}
  \State $send\_msg(next\_rank,red)$.
  \State \text{terminate}
   \ElsIf{$state = black $}
 		\State $state \gets white$.
 		\State $send\_msg(next\_rank,\&state,\&count)$.  
        \EndIf
\ElsIf{$state = red$}
 			\State $send\_msg(next\_rank,state red)$.
 			\State \text{terminate}.
    \Else
	\State $send\_msg(next\_rank,\&state,\&count)$.
	\State $count \gets 0$.
    \EndIf
\end{algorithmic}[1]
\end{algorithm}
\section{Experiments and Results}
The proposed technique is evaluated for synthetic graphs(details are as given below.) generated using ParMat. The size of these graphs are comparable to some standard graphs such Road USA, Orkut, Twitter, Coauthor. The results are average of 5 independent trials of the code. Timeout for stopping idle processor in toka1(Termination Detection technique of type 1) case is calculated using the vertices and edges in the graph.

\subsection{Experiment Setup}
\textbf{Graphs:} Most of the graphs networks are unweighted
and they are converted to weighted by assigning pseudo-random values
uniformly distributed in intervals [1, 20).
\newline
\textbf{Graph1:} It consists of  391,529 vertices and 873,775 edges. 
\newline 
\textbf{Graph2:}It is the map of the
United States roads. Each edge weight represents the distance
between a pair of vertices. It has 23,947,347 vertices
and 58,333,344 edges and maximun degree of 9.
\textbf{Graph3:} This network
has 3, 072, 441 nodes and 117, 185, 083 edges.
\newline
\textbf{Graph4:}The graph has 41.7 nodes and 1.47 billion edges.
\subsection{Results}
In this section, we will discuss the performance of our technique.Starting with the phases involved in our technique.
As discussed in section III,our algorithm works in-line with following phases: Graph Processing, Graph Partition, Edge pruning,Asynchronous SSSP and termination detection.First three phases incurs only one time cost whereas the rest two are executed for each vertex.
\newline
\indent \indent \indent We have calculated the running time for different number of processors as shown in Fig 1.It includes computation as well as communication time.Further the speedup is shown in Fig 2.To compute the work done by such algorithms it is important to have a look on MTEPS (Million Traversed Edges per Second).It also reflects the efficiency of termination detection technique used and useful to count communication time exclusively.
\section{Conclusions}
For graph1 since the size is very small and its number of vertices is comparable with number of edges which implies a lesser communication time and hence distributing among larger number of processor helps.Thus it's showing best speedup among other graphs.From fig 2.we can clearly see that the speedup for graph2 is better than three despite having large number of vertices is due to its higher number of edges than in graph3.
\includegraphics[width=\linewidth]{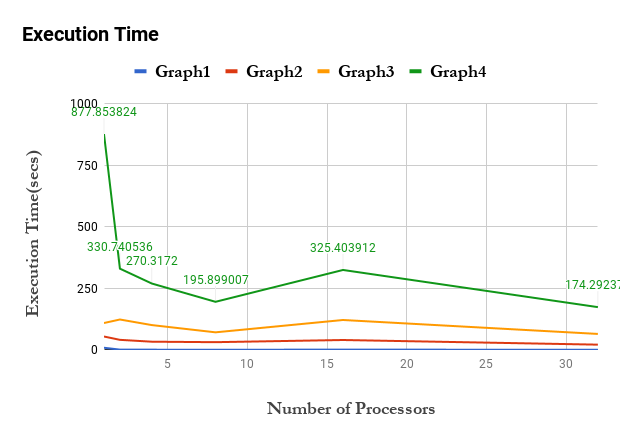}
\includegraphics[width=\linewidth]{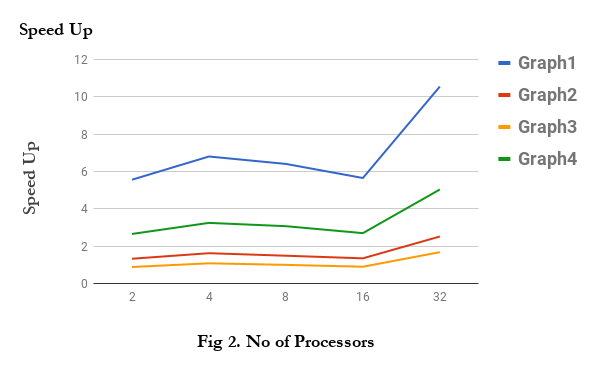}
\includegraphics[width=\linewidth]{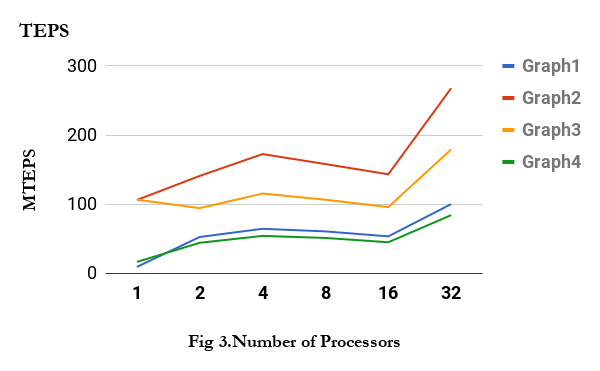}

For all the graphs except graph2 and graph3 the speedup is in sync with the size which is well reasoned in the preceding paragraph.Thus we can say that this algorithm is better suited for the graphs having moderate number of edges as compared to the vertices in it.

\section{Future Work}The asynchronous message passing of distance updates to neighboring processes can use message buffering with different push out criteria. Implementation and evaluation of such message passing scheme could be taken as future work. Termination detection also makes up major part of asynchronous algorithm. In future we can evaluate much more termination techniques tailored for SSSP problem. The given termination detection techniques are unique and can be explored with other asynchronous problems.    
\bibliographystyle{IEEEtran}
\bibliography{main}

\end{document}